\documentclass[conference]{IEEEtran}
\usepackage[utf8]{inputenc}
\usepackage{cite}
\usepackage{amsmath}
\usepackage{amsfonts}
\usepackage{amssymb}
\usepackage{mathtools}
\usepackage{amsthm}
\usepackage{graphicx}
\usepackage{epstopdf}
\usepackage{bm}
\usepackage[caption=false, font=footnotesize]{subfig}
\usepackage{xcolor}

\newtheorem{proposition}{Proposition}
\newtheorem{corollary}{Corollary}

\newcommand{\Var}{\operatorname{Var}}
\newcommand{\Cov}{\operatorname{Cov}}
\newcommand{\pr}{\operatorname{P}}

\begin{document}

\pdfoptionpdfminorversion=7

\title{A Value of Information Framework for Latent Variable Models}
\author{
    \IEEEauthorblockN{Zijing Wang, Mihai-Alin Badiu, and Justin P.~Coon}
    \IEEEauthorblockA{
        Department of Engineering Science, University of Oxford\\
        Oxford OX1 3PJ, United Kingdom\\
        e-mail: \{zijing.wang, mihai.badiu, justin.coon\}@eng.ox.ac.uk
    }
}
\maketitle

\begin{abstract}
In this paper, a general value of information (VoI) framework is formalised for latent variable models. In particular, the mutual information between the current status at the source node and the observed noisy measurements at the destination node is used to evaluate the information value, which gives the theoretical interpretation of the reduction in uncertainty in the current status given that we have measurements of the latent process. Moreover, the VoI expression for a hidden Markov model is obtained in this setting. Numerical results are provided to show the relationship between the VoI and the traditional age of information (AoI) metric, and the VoI of Markov and hidden Markov models are analysed for the particular case when the latent process is an Ornstein-Uhlenbeck process. While the contributions of this work are theoretical, the proposed VoI framework is general and useful in designing wireless systems that support timely, but noisy, status updates in the physical world.
\end{abstract}


\section{Introduction}
The freshness of data is of critical importance in wireless communication systems to support real-time management and enable precise control of entities in the real world. Stale information can be problematic. For example, in smart transportation systems, outdated safety data from autonomous vehicles may lead to severe traffic accidents. Therefore, timely status updates play a vital role in such systems.

\begin{figure}
\centering
\includegraphics[width=5.5cm]{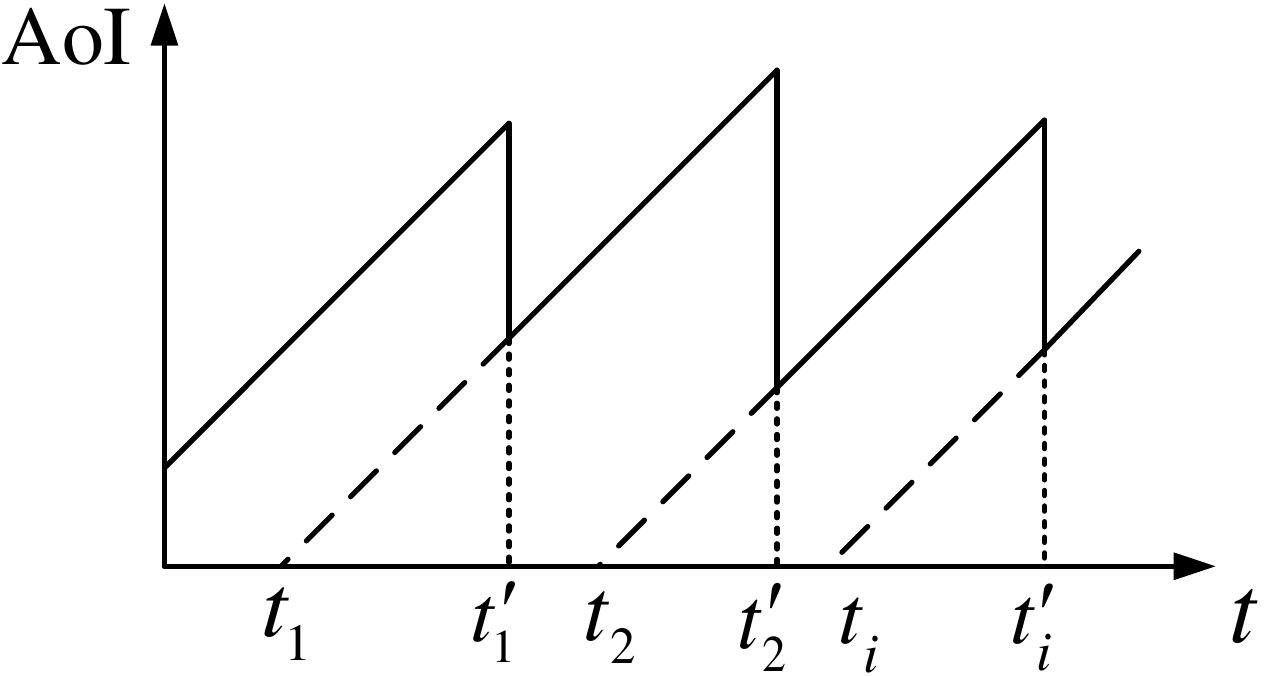}
\caption{\text {Age of information. The $i$th update is generated by the source node} \protect\\
\text{at ${t_i}$ and received by the destination node at $t'_i$.} }
\label{fig:AoI-definition}
\vspace{-0.31cm}
\end{figure}

Age of information (AoI) \cite{2012infocom} has been proposed as a new performance metric to measure the data freshness at the receiver since the last sampling at the transmitter. Specifically, AoI is defined as the time elapsed since the latest received status update was sampled. This concept is illustrated in Fig.~\ref{fig:AoI-definition}. The AoI at time $t$ can be expressed as
\begin{equation}
\label{eq:AoI-definition}
    \Delta(t)=t-u(t),
\end{equation}
where $u(t)$ is time that the latest sample received at the destination was generated. In the past few years, problems related to queueing systems \cite{queue1}\cite{queue2}, scheduling algorithms \cite{scheduling1}\cite{scheduling2}, and source coding \cite{coding1} have been widely studied with the aim of minimising AoI.

In reality, different types of data sources may change at different speeds. The notion of AoI defined in~\eqref{eq:AoI-definition} is independent of the statistical variations inherent in underlying source data. This means that the AoI metric cannot fully capture the degradation in information quality caused by the time lapse between status updates or any relevant properties the random process generated by the source might exhibit, such as how correlated it is. For example, some data sources (e.g., the engine temperature of a vehicle) change slowly over time; thus old samples may be sufficient enough to predict the future status. On the other hand, some sources (e.g., the position of a vehicle) change quickly over time, and even fresh samples with a low age may hold little useful information. It seems that old information may still have value, while new information may have less value. Therefore, it is important to take a more systematic approach to measuring the information value.

The performance of a communication system is largely affected by interference, errors, and noise. This means that the update status generated by the source node can be negatively affected, and may not be directly visible when it is received by the destination node. This motivates us to develop a general value of information (VoI) framework for latent variable models, which can be applied in many practical real-time applications.

Recently, the concept of VoI has begun to be discussed. For example, the analytic hierarchy process (AHP) was exploited in \cite{AHP1} to define VoI, and a VoI-based strategy was proposed to balance dissemination of the critical and non-critical data in vehicular networks. Furthermore, considering that the actual performance of a status update system is non-linear in the AoI, a non-linear AoI-related function was widely utilised in \cite{2016infocom}-\cite{non-linear-survey} to quantify the information value. A non-linear AoI penalty function was proposed in \cite{2016infocom}, which maps the age to a penalty function to evaluate the level of ``dissatisfaction'' related to stale information. The average AoI penalty under exponential and logarithmic functions was treated in \cite{non-linear}. A method for calculating non-linear age functions under different queueing models was proposed in \cite{TWC-EH} for energy harvesting networks. Despite these contributions, it seems that the non-linear functions in the existing work have been chosen arbitrarily without any particular theoretical basis or interpretation.

Information-theoretic VoI research has received more attention recently. For example, the estimation error was utilised as a special age penalty function in \cite{estimation1}-\cite{estimation4}. Furthermore, in \cite{SPAWC}, the mutual information function was utilised to measure the timeliness of information. In that work, data freshness was improved by optimising the sampling rate for Markov models in which the variables are assumed to be directly observable at the receiver. In practice, when we take both sampling and transmission processes into consideration, the samples at the source are latent for observation because of interference, noise, or other features that can lead to a performance degradation. Existing VoI-related works do not explicitly treat latent variable models.

In this paper, we propose a mutual information-based VoI framework for latent variable models to characterise how valuable the status updates are to the destination node. The VoI definition gives the standard interpretation of the reduction in uncertainty in the current (unobserved) status of a latent process given that we have noisy measurements. Moreover, the VoI expression is analysed for one of the most important latent variable models: the hidden Markov model (HMM) characterised by a latent Ornstein-Uhlenbeck (OU) process with noisy observations. Numerical results are provided to show the relationship between the traditional AoI metric and the proposed VoI metric in this setting. The performance of VoI for the Markov and the hidden Markov models is also discussed.

The rest of this paper is organised as follows. The VoI framework for latent variable models is formalised in Section II. The VoI for a specific, important hidden Markov model, the noisy OU process, is presented in Section III. Numerical results and discussions are provided in Section IV. Conclusions are summarised in Section V.

\section{Value of Information Formalism}
\subsection{Definition}
We consider a pair of source and destination nodes, and assume that the source node generates a sequence of time-stamped messages representing updates of the status of a random process. The messages are transmitted via a communication system to the destination node. Although ideally the receiver would receive a status update at the moment it is generated at the source, it is assumed that the communication system has limited resources, such that the message reaches the destination after some time.

Denote $\{X_t\}$ as the random process under observation, where $t$ is the time variable, which can be continuous. The message $(t_i,X_{t_i})$ is generated by the source node at arbitrary time $t_i$, and it contains this timestamp and the corresponding value $X_{t_i}$ of the process. The status updates are received by the destination node at times $t_1',t_2',\ldots$, where $t_i'>t_i$. The observations at the destination node are captured in the observed process $\{Y_t\}$, where $Y_{t_i'}$ is the observation corresponding to $X_{t_i}$. Let $n$ be the index of the most recent data received by the destination node at time $t_n'$.

The concept of VoI is defined as the mutual information between the current status of the underlying process at the source node and a sequence of observations received by the destination node before. The general definition of VoI is given as
\begin{equation}
\label{eq:general definition}
    v(t) = I(X_{t};Y_{t_1'},\ldots,Y_{t_n'}), \quad t>t'_n.
\end{equation}
Intuitively, $v(t)$ represents the reduction in uncertainty in the latent current status given that we have a collection of (possibly) noisy measurements before time $t$. This metric is appropriate for measuring the value that the past observations $\{Y_{t_i'}\}$ offer with respect to the current status of an unknown process $X_t$.

Based on the chain rule for information \cite{InfoTheo}, the general VoI expression given in \eqref{eq:general definition} for latent variable models can also be written as
\begin{equation}
\label{eq:general_voi}
v(t) = \sum\limits_{i = 1}^n {I({X_t};{Y_{t{'_i}}}|{Y_{t{'_1}}}, \cdots ,{Y_{t{'_{i - 1}}}}} ).
\end{equation}
If $\{X_t\}$ is Markov, the VoI expression can be further manipulated (cf.~sec.~\ref{sec:hmm}). Otherwise, the VoI for general latent variable models can be calculated by using the joint and marginal probability density functions (PDF) of $\{X_{t_i}\}$ and $\{Y_{t'_i}\}$.

\subsection{An Illustrative Example and a Bound}
A plethora of latent variable models exist; we do not attempt to treat all of them here.  However, it is worth considering the following simple example in an effort to elucidate the generality of the new VoI definition.  Let $\{X_t\}$ be a random process, and let $\{Y_t\}$ be the corresponding observed process, the values of which are dependent upon the latent variables.  Let $\bm{X} = {[{X_{t_1}},\cdots,{X_{t_n}}]^{\operatorname{T}}}$ and $\bm{Y} = {[Y_{t'_1},\cdots,Y_{t'_n}]^{\operatorname{T}}}$, and suppose that $X_t$ and $\bm{Y}$ are conditionally independent given the latent state vector $\bm{X}$.  Then we can write
\begin{equation}
\begin{aligned}
  v(t) &= h(X_t) - h(X_t|\bm{Y})\\
 &\le h(X_t) - h(X_t|\bm{Y},\bm{X})\\
 &= h(X_t) - h(X_t|\bm{X})\\
 &= I(X_t;\bm{X}).
\end{aligned}
\end{equation}
A similar calculation yields
\begin{equation}
    v(t) \le I(\bm{X};\bm{Y}).
\end{equation}
Combining these inequalities, we have that
\begin{equation}
\label{eq:general VoI bound}
    v(t) \le \min \{I(X_t;\bm{X}),I(\bm{X};\bm{Y})\}.
\end{equation}

If the underlying process is directly observable, then $\bm{Y} = \bm{X}$ and we immediately have that
\begin{equation}
    v(t) = I(X_t;\bm{X}).
\end{equation}
Hence, in this example, the lack of a direct route to observing $\{X_t\}$ reduces the VoI.  If, in addition to the process $\{X_t\}$ being directly observable, we also have that $\{X_t\}$ is a Markov process, then the VoI simplies further to~\cite{SPAWC}
\begin{equation}
    v(t) = I(X_t;X_{t_n}).
\end{equation}

\subsection{VoI for Hidden Markov Models}\label{sec:hmm}
\begin{figure}
\centering
\includegraphics[width=6.7cm]{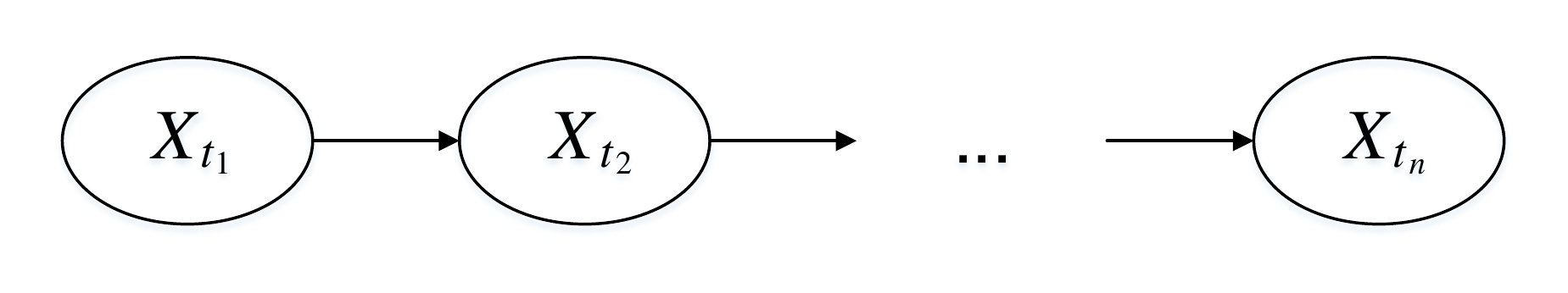}
\caption{Temporal evolution of the Markov model.}
\label{fig:MM}
\end{figure}

\begin{figure}
\centering
\includegraphics[width=7cm]{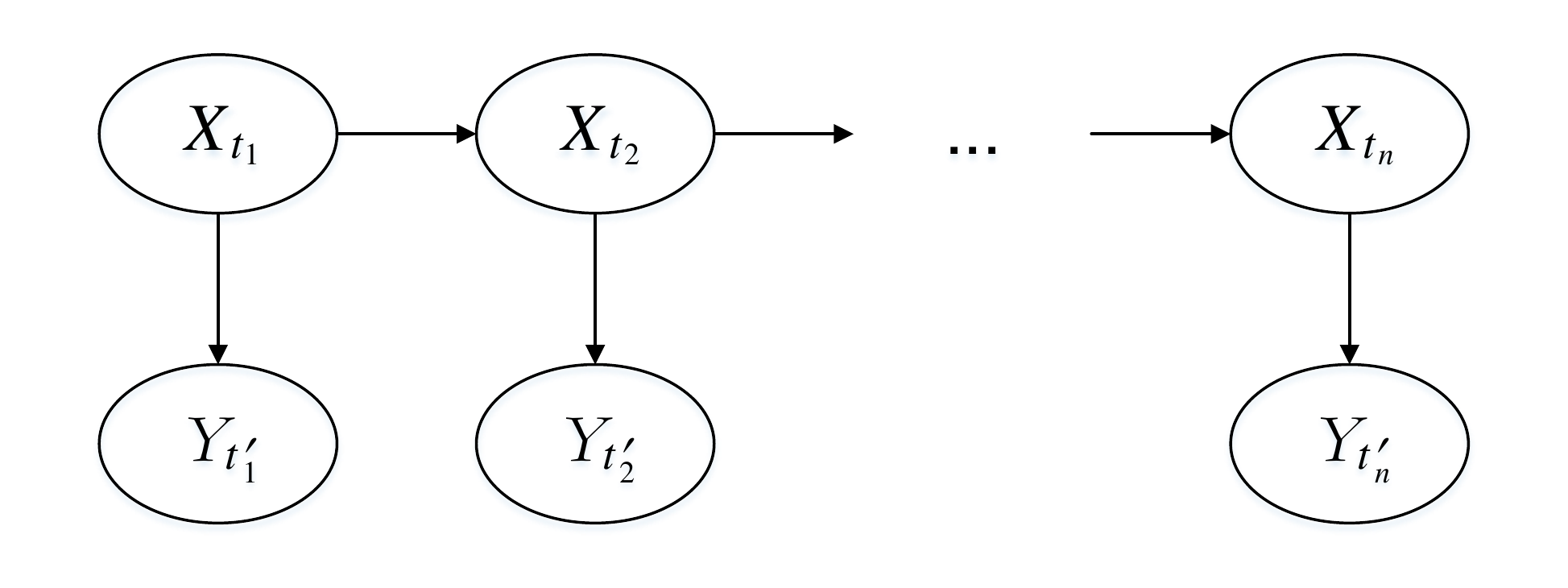}
\caption{Temporal evolution of the hidden Markov model.}
\label{fig:HMM}
\end{figure}

The hidden Markov model is an important latent variable model. Fig.~\ref{fig:MM} and Fig.~\ref{fig:HMM} illustrate the temporal evolution of the Markov and the hidden Markov models, respectively. For the Markov model in Fig.~\ref{fig:MM}, the random process can be observed directly by the receiver, i.e., $Y_{t_i'}=X_{t_i}$, for all $i=1,2,\ldots$, and the observations are Markov. While for the hidden Markov model in Fig.~\ref{fig:HMM}, $X_t$ is a Markov process, and what we receive at the destination is possibly different from the initial value, i.e., $Y_{t_i'} \not= X_{t_i}$, but where
\[
    \pr[Y_{t_i'}\in A | X_{t_1},\ldots,X_{t_i}] = \pr[Y_{t_i'}\in A | X_{t_i}]
\]
for all admissible $A$. Thus the initial samples $\{X_{t_i}\}$ are hidden at the observation node.

For the hidden Markov model, the mutual information $I(X_t;\bm{X})$ can be expressed as
\begin{equation}
    I({X_t};\bm{X}) = \sum\limits_{i = 1}^n {I({X_t};{X_{{t_i}}}|{X_{{t_{i - 1}}}})}
\end{equation}
based on the chain rule. Similarly, we can show that
\begin{equation}
    \begin{aligned}
        I(\bm{X};\bm{Y})
            &= \sum\limits_{i = 1}^n {I(\bm{Y};{X_{{t_i}}}|{X_{{t_{i - 1}}}})} \\
            &= \sum\limits_{i = 1}^n I(X_{{t_i}};Y_{t_i'}|{X_{{t_{i - 1}}}}).
    \end{aligned}
\end{equation}
Therefore, the VoI bound for the general latent variable model in~\eqref{eq:general VoI bound} can be rewritten as
\begin{equation}
\label{eq:HMM VoI bound}
    v(t) \le \min \left\{\sum\limits_{i = 1}^n {I({X_t};{X_{{t_i}}}|{X_{{t_{i - 1}}}})};\sum\limits_{i = 1}^n I(X_{{t_i}};Y_{t_i'}|{X_{{t_{i - 1}}}})\right\}.
\end{equation}

\section{VoI for a Noisy OU Process}\label{sec:ou}

\subsection{Noisy OU Process Model}
As an important example of how the VoI framework can be applied, we consider the case of a noisy Ornstein–Uhlenbeck (OU) process.  In this case, the random process $X_t$ at the source node satisfies the stochastic differential equation (SDE)
\begin{equation}
\label{eq:OU process}
    \operatorname{d}\!X_t = \kappa (\theta-X_t) \operatorname{d}\!t + \sigma\operatorname{d}\!W_t
\end{equation}
where $\{W_t\}$ is standard Brownian motion, $\kappa$ is the rate of mean reversion, $\theta$ is the long-term mean, and $\sigma$ is the volatility of the random fluctuation\footnote{In practice, such a model can be used to represent the position of an autonomous device, such as an unmanned aerial vehicle (UAV) anchored to a point $\theta$ but experiencing positional disturbances due to wind.}. The OU process described by this SDE is stationary, Markov, and Gaussian.  For any $t$, the variable $X_t$ is normally distributed with mean
\begin{equation}
    \mathbb{E}[{X_t}] = \theta  + ({X_0} - \theta ){e^{ - \kappa t}}
\end{equation}
and variance
\begin{equation}
\label{eq:variance xt}
    \Var[X_t] = \frac{\sigma^2}{2\kappa}\left(1-e^{-2\kappa t}\right).
\end{equation}
$X_t$ conditioned on $X_s$ is also Gaussian with mean
\begin{equation}
   \mathbb{E}[{X_t}|{X_s}] = \theta  + ({X_s} - \theta ){e^{ - \kappa (t - s)}}
\end{equation}
and variance
\begin{equation}
\label{eq:variance xt xtn}
    \Var[{X_t}|{X_s}] = \frac{\sigma^2}{2\kappa}\left(1-e^{-2\kappa (t-s)}\right).
\end{equation}
The covariance matrix of $\bm{X}$ can be expressed as
\begin{equation}
\label{eq:cov x}
    {\mathbf{\Sigma} _{\bm{X}}} = {\left[ {\begin{array}{*{20}{c}}
{\Cov[{X_{{t_1}}},{X_{{t_1}}}]}& \cdots &{\Cov[{X_{{t_1}}},{X_{{t_n}}}]}\\
 \vdots & \ddots & \vdots \\
{\Cov[{X_{{t_n}}},{X_{{t_1}}}]}& \cdots &{\Cov[{X_{{t_n}}},{X_{{t_n}}}]}
\end{array}} \right]}
\end{equation}
where
\begin{equation}\label{eq:cov xts}
    \Cov[{X_t},{X_s}] = \frac{{{\sigma ^2}}}{{2\kappa }}({e^{ - \kappa |t - s|}} - {e^{ - \kappa (t + s)}}).
\end{equation}

We assume that the status updates are sampled at arbitrary times $t_1,t_2,\ldots$ and arrive at the destination node at times $t_1',t_2',\ldots$, where $t_i'>t_i$. We assume $\{X_t\}$ is a latent process that is observed through an additive noise channel.  Hence, this noisy OU process constitutes a hidden Markov model with observations defined by the equation
\begin{equation}
     Y_{t_i'}=X_{t_i}+N_{t'_i}
\end{equation}
where $\{N_{t'_i}\}$ is a noise process that is anchored at time $t'_i$ and which evolves with time. Practically, $N_{t'_i}$ can represent a measurement or transmission error that corrupts the update $X_{t_i}$. Assume that the noise process $\{N_{t'_i}\}$ is a Gaussian process with zero mean and variance $\Var[N_{t'_i}]$. Let the vector $\bm{X} = {[{X_{t_1}},\cdots,{X_{t_n}}]^{\operatorname{T}}}$ capture the set of status updates at the source node, and let $\bm{Y} = {[Y_{t'_1},\cdots,Y_{t'_n}]^{\operatorname{T}}}$ denote the corresponding observations received at the destination node.  Similarly, we collect the associated noise samples in the vector $\bm{N} = {[N_{t'_1},\cdots,{N_{t'_n}}]^{\operatorname{T}}}$, the covariance matrix of which is given by
\begin{equation}
    \label{eq:cov n}
        (\bm{\Sigma}_{\bm{N}})_{ij} = \left\{ {\begin{array}{*{20}{l}}
    {\Var[{N_{{t'_i}}}]}, & {i = j}\\
    0, & {i \ne j}
    \end{array}} \right.
\end{equation}
Therefore, the observations of the noisy OU process are collectively represented by
\begin{equation}
\label{eq:OU hidden mapping}
     \bm{Y}=\bm{X}+\bm{N}.
\end{equation}

\subsection{VoI for the Noisy OU Process}
Based on the model described above, we can state the following main result of this section.
\begin{proposition}\label{prop:1}
    Let $\mathbf{A} = \mathbf{\Sigma}^{ - 1}_{\bm{X}} + \mathbf{\Sigma} ^{ - 1}_{\bm{N}}$, and let $\mathbf A_{ij}$ denote the $(n-1)\times (n-1)$ matrix constructed by removing the $i$th row and the $j$th column of $\mathbf A$.  The VoI for the noisy OU process defined above can be written as
    \begin{multline}
    \label{eq:voi expresseion n dimension}
          v(t) =  \frac{1}{2}\log \bigg(\frac{{1 - {e^{ - 2\kappa t}}}}{{1 - {e^{ - 2\kappa (t - {t_n})}}}}\bigg) \\
          - \frac{1}{2}\log \bigg(1 + \frac{{{2\kappa }}}{{\sigma^2 \left(e^{2\kappa (t-t_n)} - 1\right) }} \frac{\det(\mathbf{A}_{nn})}{\det(\mathbf{A})}\bigg).
    \end{multline}
\end{proposition}
\begin{IEEEproof}
    See the appendix.
\end{IEEEproof}

It is easy to show that the first term in~\eqref{eq:voi expresseion n dimension} represents the VoI for the non-noisy Markov OU process $\{X_t\}$. Hence, the second term quantifies a ``correction''  to the VoI of the latent process that arises due to the indirect observation of the process through the noisy channel.  Note that both $\mathbf{A}$ and $\mathbf{A}_{nn}$ are positive semidefinite.  As a result, the second logarithm in~\eqref{eq:voi expresseion n dimension} is non-negative, which verifies the reduction in VoI (relative to the Markov model) promised by~\eqref{eq:general VoI bound}.
\subsection{Results for a Single Observation}
The result given in Proposition~\ref{prop:1} is general. To explore this result further, we consider the special case where one may wish to know how much value the most recently received observation (at time $t'_n$) contains about the status of a process at time $t$. In this case, the VoI given by~\eqref{eq:general definition} can be simplified to
\begin{equation}
\label{eq:simplified definition}
    v(t) = I(X_{t};Y_{t_n'}), \quad t>t'_n.
\end{equation}
This VoI metric can be calculated by replacing the $n$-dimensional vector $\bm{Y}$ with the single variable $Y_{t'_n}$ in~\eqref{eq:voi expresseion n dimension}, which leads to the following corollary.

\begin{corollary}\label{cor:1}
The VoI for the noisy OU process with a single observation is given by
\begin{multline}
\label{eq:general voi 1 dimension}
    v(t) =  \frac{1}{2}\log \bigg(\frac{{1 - {e^{ - 2\kappa t}}}}{{1 - {e^{ - 2\kappa (t - {t_n})}}}}\bigg) \\
          - \frac{1}{2}\log \left(1 +  \frac{1-e^{-2\kappa t_n}}{(1+\gamma_{n})(e^{2\kappa(t-t_n)} - 1)}\right)
\end{multline}
where $\gamma_{n} = {\Var[{X_{{t_n}}}]}/{\Var[N_{t'_n}]}$. Furthermore, as $t_n\to\infty$, we have
\begin{equation}
\label{eq:voi 1 dimension}
    v(t) \sim  \frac{1}{2}\log \left(1 + \frac{1}{\frac{1+\gamma_{n}}{\gamma_{n}}e^{2\kappa (t - t_n)} -1}\right).
\end{equation}
\end{corollary}
\begin{IEEEproof}
    The corollary follows directly from Proposition~\ref{prop:1} where $\det(\mathbf{A}_{nn})\coloneqq 1$.
\end{IEEEproof}
This shows that for fixed $t_n$, as time $t$ increases, the VoI will decrease like $O(e^{-2\kappa t})$ until a new update is received.  An update causes a corresponding reset of $v(t)$. This is somewhat similar to the AoI, $\Delta(t)$, which is equal to $t'_n-t_n$ at the moment the $n$th update arrives and then increases with unit slope until the next update comes.

Note that the parameter $\{\gamma_{i}\}$ for VoI in the hidden Markov model can evolve with time for different updates, and this can be compared to the signal-to-noise ratio (SNR) in wireless systems. The parameter $\gamma_{n}$ is able to reflect the channel condition between the source and the destination from time $t_n$ to $t'_n$. 
For a single observation, Corollary~\ref{cor:1} shows that the VoI for the noisy OU process depends on the parameter $\gamma_{n}$, which provides a comparison between the randomness inherent in the OU process and the noise in the communication channel.  When $\gamma_{n}$ is large, the OU randomness dominates, and we expect the noisy channel to play a small role in the calculation of the VoI.  On the other hand, when $\gamma_{n}$ is small, the noisy channel catastrophically corrupts the observation.  In general, $\{\gamma_i\}$ is, itself, a (nonstationary) stochastic process that reflects the channel condition between the source and the destination.  This means that the proposed VoI metric can capture both temporal and physical properties of the system, whereas the traditional AoI metric can only reflect temporal properties.  The relationship between the VoI of the noisy OU process and $\gamma_{n}$ is formalised in the following corollary.

\begin{corollary}\label{cor:2}
As $\gamma_{n} \to \infty$, the VoI of the noisy OU process converges to the VoI of the underlying process
\begin{equation}
\label{eq: voi for MM}
    v(t)\to \frac{1}{2}\log \left(\frac{1 - e^{- 2\kappa t}}{1 - e^{- 2\kappa (t - t_n)}}\right).
\end{equation}

As $\gamma_{n} \to 0$, $X_{t}$ and $Y_{t_n'}$ become independent, and $v(t)\to 0$.
\end{corollary}
\begin{IEEEproof}
    This corollary can be verified formally by letting $\gamma_{n} \to \infty$ and $\gamma_{n}\to 0$ in Corollary~\ref{cor:1}.
\end{IEEEproof}

Note that these results give extreme cases where the bound given in~\eqref{eq:general VoI bound} is met with equality.  Indeed, in the case of the first part of Corollary~\ref{cor:2}, we have that $v(t) = I(X_t;X_{t_n})$, whereas for the second part, $v(t) = I(X_{t_n};Y_{t_n'})$.  More generally, the upper bound of VoI for the noisy OU process satisfies
\begingroup
\renewcommand{\arraystretch}{2}
\begin{equation}\label{eq:OU VoI bound}
    v(t) \leq \left\{
    \begin{array}{ll}
        v_\text{OU}(t), & \gamma_{n} \geq \frac{e^{-2\kappa (t - t_n)} - e^{-2\kappa t}}{1 - e^{-2\kappa (t - t_n)}} \\
        v_\text{AGN}, & \gamma_{n} < \frac{e^{-2\kappa (t - t_n)} - e^{-2\kappa t}}{1 - e^{-2\kappa (t - t_n)}}
    \end{array}
    \right.
\end{equation}
\endgroup
where $v_\text{OU}(t)$ is the VoI of the latent (Markov) OU process given by the right-hand side of~\eqref{eq: voi for MM} and $v_\text{AGN} = (1/2) \log(1 + \gamma_{n})$ is the mutual information corresponding to the additive Gaussian noise channel $I(X_t;Y_{t_n'})$.  Eq.~\eqref{eq:OU VoI bound} captures the point (in terms of $\gamma_{n}$) at which the bound in~\eqref{eq:general VoI bound} transitions from the latent process to the noisy process.

\section{Performance Evaluation}
In this section, numerical results are presented to explore how VoI relates to AoI and to ascertain the difference in VoI for noisy and directly observed OU processes.

\subsection {Simulation Setup}
We consider a noisy OU model with one source and one destination. A sequence of messages representing the status updates of the underlying OU process are generated by the source node, and then transmitted to the destination node. In the simulation, we assume that the source generates $10$ update packets and delivers the samples to the destination at the random times. The sampling and receiving times corresponding to each update are $\{1,$ $3,$ $5,$ $7,$ $9,$ $11,$ $13,$ $15,$ $17,$ $19\}$ and $\{2,$ $4.2,$ $6.5,$ $8,$ $10.7,$ $12.1,$ $14.7,$ $16.2,$ $18.3,$ $20.5\}$, respectively. Parameter $\sigma$ is $10$.
\begin{figure}
\centering
\includegraphics[width=7cm]{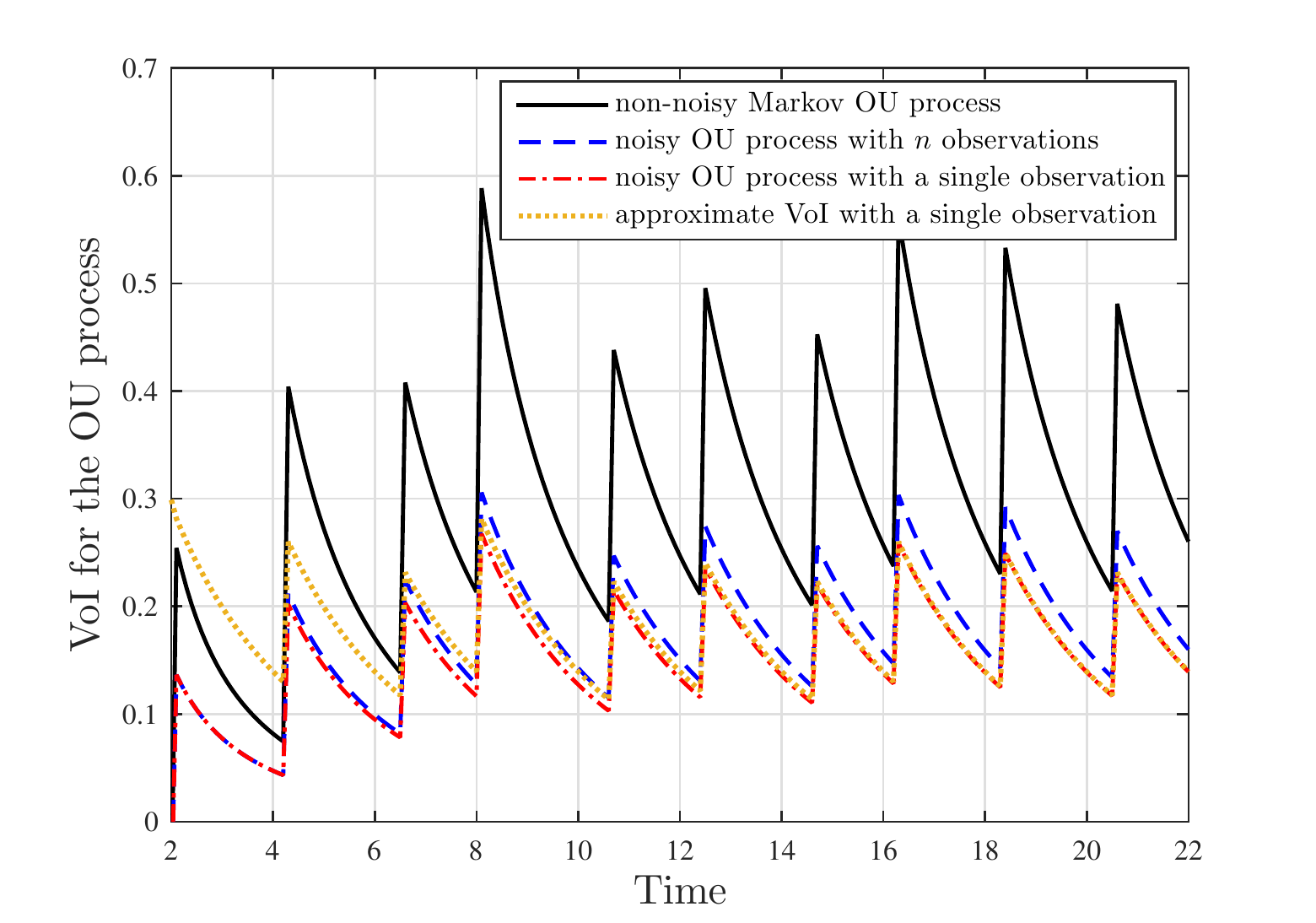}
\caption{VoI in the non-noisy Markov OU process and the noisy OU process.}
\label{fig:MM vs HMM}
\end{figure}

\begin{figure}
\centering
\includegraphics[width=7cm]{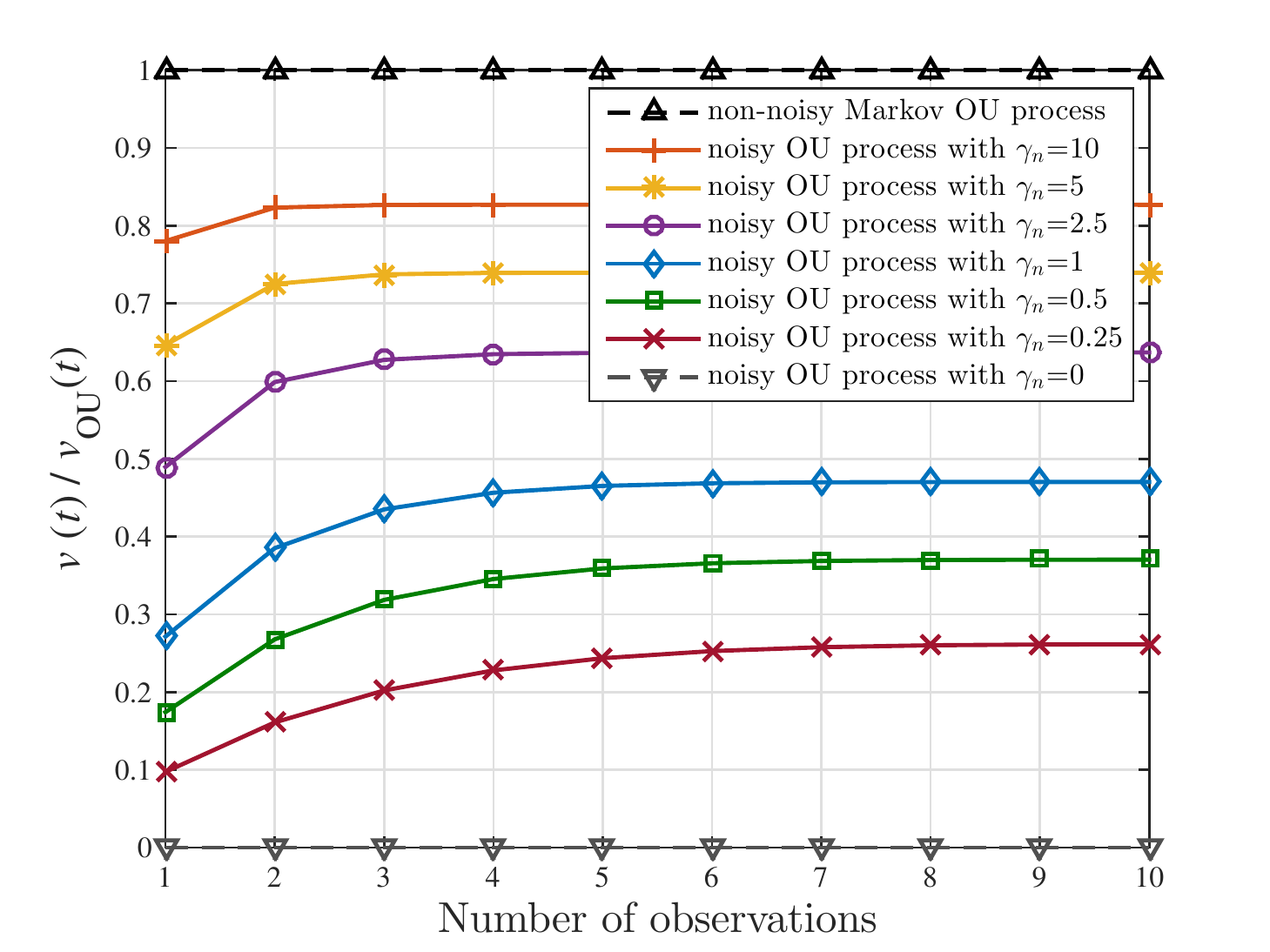}
\caption{VoI in noisy OU processes for different number of observations and parameter $\gamma_n$ at $t=21$.}
\label{fig:num of observations}
\end{figure}

\subsection {Results and Analysis}
Fig.~\ref{fig:MM vs HMM} shows the VoI for the directly observed OU process and the noisy OU process for different numbers of observations. Here, $\{\gamma_{i}\}$ is $1.5$, and $\kappa$ is $0.15$. All of the observations $\{Y_{t'_{1}}, \cdots, Y_{t'_n}\}$ received before time $t$ ($t'_n \leq t < t'_{n+1}$) are used for the results labelled ``noisy OU process with $n$ observations''. Only the most recently received observation is used for the results labelled ``noisy OU process with a single observation''. The VoI for the OU model is the first term in~\eqref{eq:voi expresseion n dimension}. The gap between the VoI of the OU model and its noisy counterpart with $n$ observations represents the second term in~\eqref{eq:voi expresseion n dimension}, which quantifies the correction to the VoI of the latent process. Similarly, the gap between the curves for the OU process and the noisy OU process with a single observation illustrates the second term in~\eqref{eq:general voi 1 dimension}. The gap between the curves for the two noisy OU processes increases with time, which illustrates that more observations gives more information about the current status of the random process. This figure verifies the reduction in VoI for the latent OU model given in Proposition \ref{prop:1} and Corollary \ref{cor:1}.  Furthermore, the approximate VoI with a single observation for the noisy OU process when $t_n \to \infty$ is also given in Fig.~\ref{fig:MM vs HMM}. The gap between the approximate VoI and the VoI corresponding to a single observation narrows with time, as expected.
\begin{figure}
\centering
\includegraphics[width=7cm]{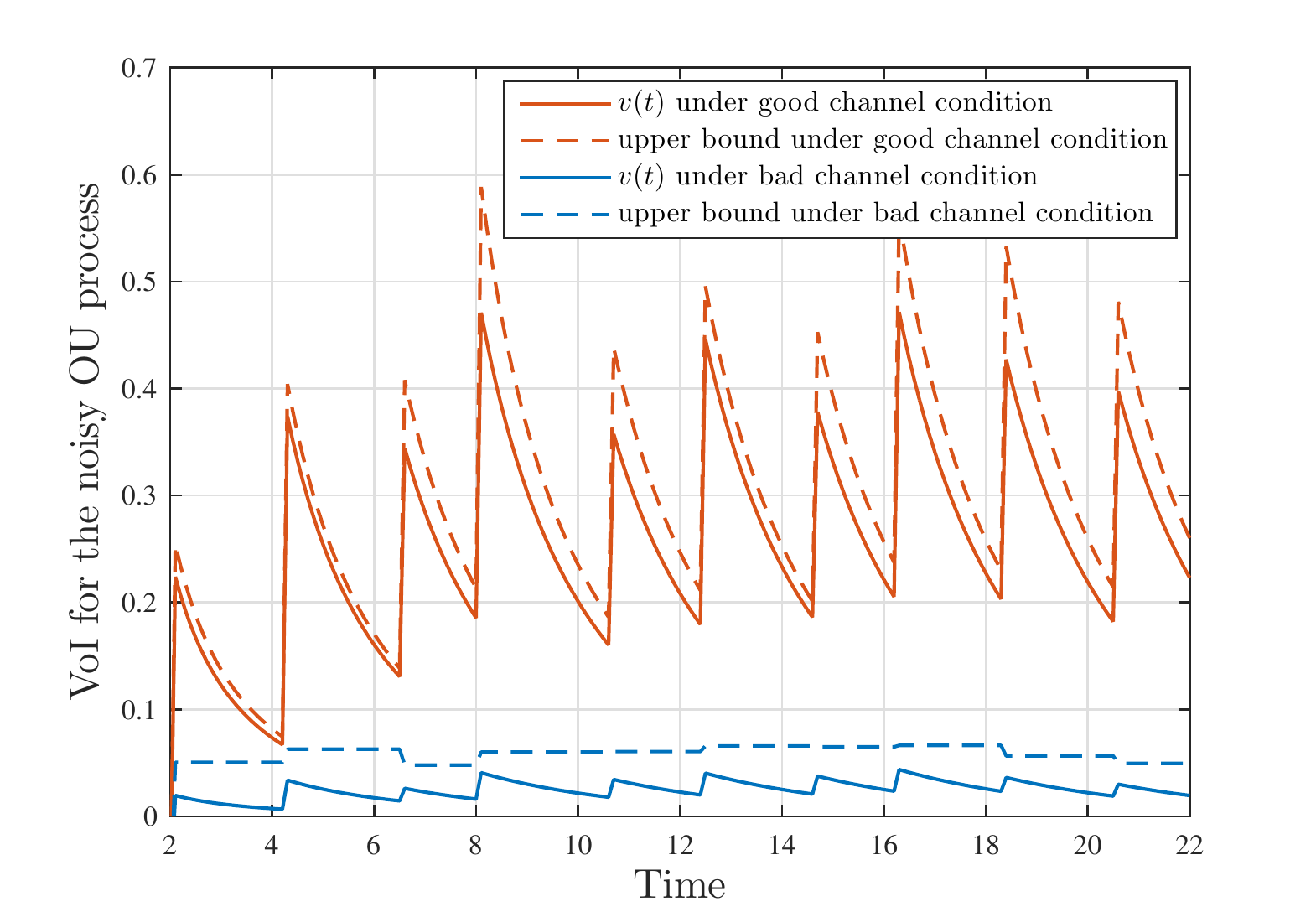}
\caption{Upper bound of VoI under different channel conditions.}
\label{fig:upper bound}
\end{figure}

\begin{figure}
\centering
\includegraphics[width=7cm]{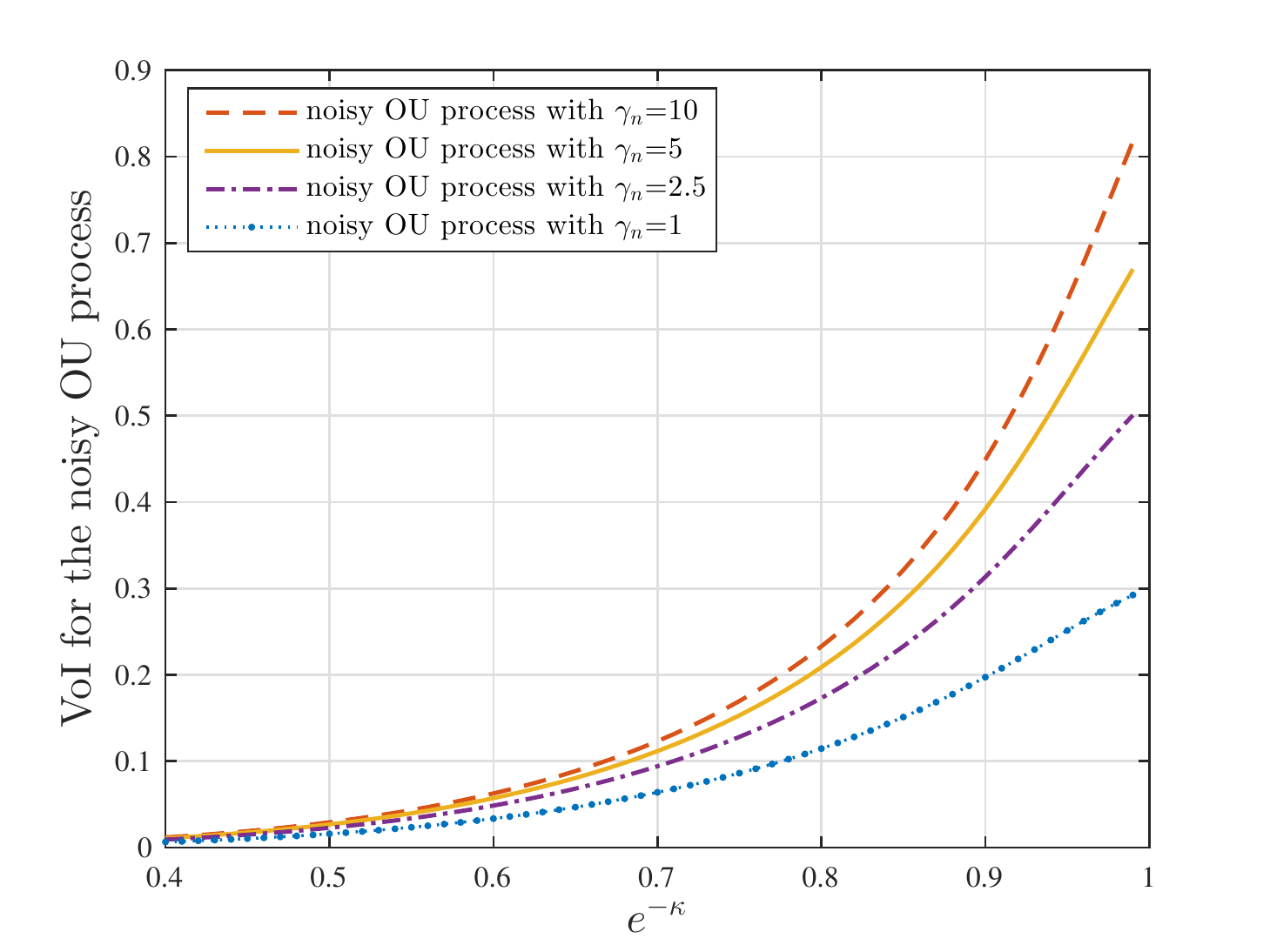}
\caption{VoI in noisy OU processes for different $\kappa$ at $t=21$.}
\label{fig:value-kappa}
\end{figure}

Fig.~\ref{fig:num of observations} shows how the VoI varies with the number of measurements $k$ for different values of $\gamma_n$. Here, $\kappa$ is $0.025$. The observations $\{Y_{t'_{n-k+1}}, \cdots, Y_{t'_n}\}$ are used for the noisy OU process. The horizontal axis is the number of observations. The vertical axis represents the ratio of $v(t)$ to $v_\text{OU}(t)$, where $v_\text{OU}(t)$ is the VoI in underlying OU process. This result illustrates that the VoI increases with the number of observations, converging to a constant as more past observations are used. Moreover, $v(t)$ approaches $v_\text{OU}(t)$ as $\gamma_n$ increases, and $v(t)$ approaches $0$ as $\gamma_n$ decreases (see~Corollary \ref{cor:2}).

The upper bound of VoI given in~\eqref{eq:general VoI bound} is plotted in Fig.~\ref{fig:upper bound} along with the VoI given in Proposition \ref{prop:1}. Here, $\kappa$ is $0.15$, and $\{\gamma_i\}$ was generated randomly to represent good and bad channel conditions. For the case where the channel conditions are good, the set $\{9.62,$ $19.01,$ $8.29,$ $7.47,$ $8,$ $15.29,$ $8.16,$ $9.52,$ $7.08,$ $7.89\}$ was generated; for the case of bad channel conditions, the set $\{0.11,$ $0.13,$ $0.1,$ $0.13,$ $0.13,$ $0.14,$ $0.14,$ $0.14,$ $0.12,$ $0.1\}$ was generated. This figure illustrates that when the noise induced by the channel is low, OU randomness is dominant, and the upper bound is the VoI for the underlying process. When $\gamma_i$ is small, we observe the alternative result.  Interestingly, we see the bounds are reasonably tight for the OU example.

Fig.~\ref{fig:value-kappa} shows the VoI (with a single observation) for the noisy OU process with different values of $\kappa$. The mean reversion parameter $\kappa$ captures the correlation of the latent random process. This figure illustrates that the value of highly correlated samples is larger, compared with the less correlated samples, which shows that  ``old" samples from the highly correlated source may still offer value.


\section{Conclusions}
In this paper, a general value of information framework for latent variable models was formalised. The concept of VoI was defined here as the mutual information between the current status of a latent random process and the sequence of past noisy measurements. This VoI metric gives the interpretation of the reduction in uncertainty in the current status given that we have noisy observations, and it is appropriate for measuring how valuable status updates from a source are at a destination node. The VoI expression for a typical latent variable model (a noisy OU process) was obtained. Comparing with the traditional AoI metric, the proposed VoI framework captures not only the time evolution of the random process, but also the correlation of updates at the source and noise in the transmission environment.

\section*{Appendix}
Since $(\bm{Y}^{\operatorname{T}},X_t)$ is multivariate Gaussian distributed, it follows from the relation $I(X_t;\bm{Y}) = h(X_t) + h(\bm{Y}) - h(X_t,\bm{Y})$ and the definition given in~\eqref{eq:general definition} that
\begin{equation}
\label{eq:genernal voi with covariance}
   v(t) = \frac{1}{2}\log \frac{{\Var[X_t]\det({\mathbf{\Sigma} _{\bm{Y}}})}}{\det({\mathbf{\Sigma} _{\bm{Y},X_t}})}.
\end{equation}
Here, ${\mathbf{\Sigma} _{\bm{Y}}}$ and $\mathbf{\Sigma} _{\bm{Y},X_t}$ are the covariance matrices of ${\bm{Y}}$ and $({\bm{Y}}^{\operatorname{T}},X_t)^{\operatorname{T}}$, respectively.

As ${\bm{X}}$ and ${\bm{N}}$ are independent, the covariance matrix ${\mathbf{\Sigma} _{\bm{Y}}}$ is given as
\begin{equation}
\label{eq:cov y}
   {\mathbf{\Sigma} _{\bm{Y}}} = {\mathbf{\Sigma} _{\bm{X}}} + {\mathbf{\Sigma} _{\bm{N}}}.
\end{equation}
$\det(\mathbf{\Sigma} _{\bm{Y},X_t})$ can be obtained by the PDF of $(\bm{Y}^{\operatorname{T}},X_t)$, and the PDF of $(\bm{Y}^{\operatorname{T}},X_t)$ can be obtained by marginalising the joint PDF of $(\bm{Y}^{\operatorname{T}},X_t,\bm{X}^{\operatorname{T}})$ over $\bm{X}^{\operatorname{T}}$, i.e.,
\begin{equation}
\label{eq:cov y xt}
    \det({\mathbf{\Sigma} _{\bm{Y},X_t}}) =\Var[{X_t}|{X_{{t_n}}}]  \det\bigg({\mathbf{\Sigma} _{\bm{N}}} + {\mathbf{\Sigma} _{\bm{X}}} + \frac{{{\mathbf{\Sigma} _{\bm{X}}}\mathbf{v}\mathbf{v}^{\operatorname{T}}{\mathbf{\Sigma} _{\bm{N}}}}}{{\Var[{X_t}|{X_{{t_n}}}]}}\bigg)
\end{equation}
where vector $\mathbf{v} = [0, \cdots ,0,{e^{ - \kappa (t - {t_n})}}]^{\operatorname{T}}$.

Substituting~\eqref{eq:cov y} and~\eqref{eq:cov y xt} into~\eqref{eq:genernal voi with covariance}, the VoI for the noisy OU process can be expressed as
\begin{equation}
v(t) =  \frac{1}{2}\log \left(\frac{{\Var[{X_t}]}}{{\Var[{X_t}|{X_{{t_n}}}]}} \frac{{\det({\mathbf{\Sigma} _{\bm{N}}} + {\mathbf{\Sigma} _{\bm{X}}})}}{{\det({\mathbf{\Sigma} _{\bm{N}}} + {\bm{\Sigma} _{\bm{X}}} + \frac{{{\mathbf{\Sigma} _{\bm{X}}}\mathbf{v}\mathbf{v}^{\operatorname{T}}{\mathbf{\Sigma} _{\bm{N}}}}}{{\Var[{X_t}|{X_{{t_n}}}]}})}}\right).
\end{equation}
By applying matrix determinant lemma, this expression can be further simplified to
\begin{multline}
    v(t) = \frac{1}{2}\log \bigg(\frac{{1 - {e^{ - 2\kappa t}}}}{{1 - {e^{ - 2\kappa (t - {t_n})}}}}\bigg) \\
    - \frac{1}{2}\log \bigg(1 + \frac{{{2\kappa }}}{{\sigma^2 \left(e^{2\kappa (t-t_n)} - 1\right) }} \frac{\det(\mathbf{A}_{nn})}{\det(\mathbf{A})}\bigg).
\end{multline}

\end{document}